# EPR studies of the 105 K phase transition of SrTiO$_3$ with the non-cubic Fe$^{5+}$ as local probe and a reinterpretation of other Fe$^{5+}$ centres

Th. W. Kool

Van 't Hoff Institute for Molecular Sciences, University of Amsterdam
the Netherlands

January 2012

**Abstract**

The 105 K second-order displacive phase transition of SrTiO$_3$ has been studied with the help of Electron Paramagnetic Resonance. The photochromic non-cubic Fe$^{5+}$ centre is used as local probe. Critical phenomena, characterized by an exponent $\beta = 1/3$, are presented. Line broadening effects are interpreted as stemming from time-dependent fluctuations near the phase transition point. The data are consistent with those reported earlier on the Fe$^{3+}$-V$_O$ pair centre, indicating cooperative effects in the crystal. Also a model for the non-cubic Fe$^{5+}$ is proposed, i.e., the ion is substitutional for Ti$^{4+}$ with an empty adjacent expanded octahedron. Other Fe$^{5+}$ centres in SrTiO$_3$ and BaTiO$_3$ are reviewed and reinterpreted.

**Introduction**

Displacive phase transitions were analyzed almost four decades ago with the help of classical theories such as the Landau theory[1] or with microscopic theories using the mean field approximation.[2] More specifically, a typical temperature behaviour is predicted for the generalized susceptibility $\chi$, the order parameter and the specific heat $c_p$ of the form $(-\varepsilon)^x$, where $\varepsilon = (T - T_c)/T_c$.

Values for the exponent $x$ are:

$x = \beta = \frac{1}{2}$ for the order parameter,

$x = \gamma = -1$ for the susceptibility and

$x = 0$ for the specific heat, indicating a jump at $T_c$, where $T_c$ is the critical temperature of the phase transition.

Although classical theories are adequate for temperatures well outside the phase transition point, deviations or critical behaviour from these theories occur near $T_c$. These phenomena arise from correlated fluctuations of the order parameter and become important when the length of the correlated fluctuations exceeds the range of forces. An explanation of these effects and the EPR technique used is reviewed by the 1987 Nobel laureate K. Alex Müller and J.C. Fayet.[3] This article can also be found in Chapter VII of the book: *Properties of Perovskites and other oxides* by K. Alex Müller and Tom W. Kool.[4]

Experimentally, critical behaviour of SrTiO$_3$ (STO) near the 105 K phase transition was verified by means of the Fe$^{3+}$ and Fe$^{3+}$-V$_O$ impurity centres,[5] both are substitutional for Ti$^{4+}$. In this paper we present EPR results of the 105 K second-order displacive phase transition in STO, where the non-cubic photochromic Fe$^{5+}$ is used as local probe. This impurity centre is a $d^3$ ($S = 3/2$) system, substituting for Ti$^{4+}$ and is octahedrally surrounded by a cage of oxygen ions in the presence of a moderate axial field for $T > T_c$. For $T < T_c$ a weak orthorhombic perturbation due to the phase transition is added. This centre has been analyzed before by Kool *et al.*[6]

The study of phase transitions by means of EPR and the use of *different impurities* situated at the *same* site provides more evidence of the cooperative behaviour in the crystal. The non-cubic Fe$^{5+}$ centre in STO is adequate because of the large anisotropy of the resonance lines, ranging from $g \approx 2 - 4$ and the very accurate measurements of the rotational order parameter $\varphi(T)$.

**The non-cubic Fe$^{5+}$ centre**

For axially distorted (tetragonal or trigonal) octahedrally surrounded $d^3$ spin systems the following spin-Hamiltonian is used:[7-10]

$$\mathcal{H} = S \cdot \bar{D} \cdot S + \mu_B H \cdot \bar{g} \cdot S \tag{1}$$

The first term represents the zero-field splitting and the latter the Zeeman interaction. For systems with $|D| \gg h\nu$, with $\nu$ the frequency of a typical EPR experiment, the first term is taken as zero order Hamiltonian $\mathcal{H}_0$ and the Zeeman splitting is treated as a perturbation $\mathcal{H}_1$. For values $h\nu/|2D| \geq 0.25$ one has to proceed with exact numerical computer calculations.[11] If the zero-field splitting $|2D|$ is much larger than the Zeeman term, only one EPR transition within the Kramers doublet with $M_S = |\pm 1/2\rangle$ levels is observed giving a typical $g^{eff}$ EPR spectrum for $S = 3/2$ systems ranging from $g^{eff} \approx 2 - 4$. The EPR spectrum of the non-cubic Fe$^{5+}$ centre shows these lines too.[6] The non-cubic Fe$^{5+}$ centre has the following $g$ and $D$ values: $g_\parallel = 2.0132$, $g_\perp = 2.0116$ and $|2D| = 0.541$ cm$^{-1}$ at 115 K.[4,6] From depopulation measurements at helium temperatures it could be concluded that the sign of $D$ is negative.

Below the phase transition ($T < T_c$) an extra weak orthorhombic perturbation in the spin-Hamiltonian has to be added:

$$E(S_x^2 - S_y^2). \tag{2}$$

For $|D| \geq h\nu$ and $|E| \leq h\nu$, general angular expressions for $d^3$ ($S = 3/2$) systems were derived.[12]

At 77 K, well below $T_c$, $|2D| = 0.551$ cm$^{-1}$ and $|E| = 0.529 \times 10^{-3}$ cm$^{-1}$, indicating that $D$ is a little temperature dependent. The rhombic parameter $E$ is temperature dependent and is proportional to $\varphi^2$, with φ the intrinsic rotation angle of STO consisting of alternating rotations of neighbouring oxygen octahedra below $T_c$.[4,6]

The rotation angle $\varphi^*$ of the non-cubic $Fe^{5+}$ is larger than the intrinsic one φ. The relative large value for $\varphi^*$ is interpreted as follows. As can been seen in Fig. 1 the octahedron adjacent to the $Fe^{5+}$ ion has been expanded. As a consequence of this expansion the rotation angle $\varphi^*$ becomes larger.

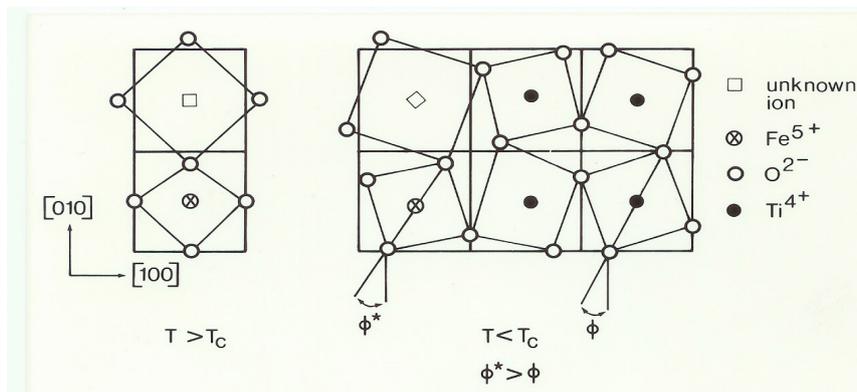

*Fig.1. Rotation of the oxygen octahedra around the [001] axis. φ is the intrinsic rotation angle of the crystal, while $\varphi^*$ is the rotation angle of the non-cubic $Fe^{5+}$ centre, which has an expanded adjacent oxygen octahedron due to a $Ti^{4+}$ vacancy.*

This expansion can be caused by a $Ti^{4+}$ vacancy or a nearby impurity substitutional for the $Ti^{4+}$ ion. Simple calculations at 77 K showed that with the given intrinsic angle φ = 1.53° and $\varphi^*$ = 1.75° of the non-cubic $Fe^{5+}$ and the distance of the lattice constant (above $T_c$) a = 3.9 Å, the expansion must be equal to 0.30 Å. It follows then that the unknown ion possesses a radius r = 0.94 Å, knowing that $r(Ti^{4+})$ = 0.64 Å.[13]

The impurity concentration (in ppm) in the investigated crystal obtained by spectrochemical analyses is as follows (see the Table):

**Table**

| Fe 18 | B<10 |
|---|---|
| Mo 2 | Si 500-2000 |
| Pb 500-1000 | Ni 10-50 |
| Sn 200-1000 | Al<10 |

*Impurity concentration in ppm.*

None of the above mentioned impurities and their respective ions fit into the expanded cage.[13] Also $Ti^{2+}$, with $r(Ti^{2+})$ = 0.80Å, and $Sr^{2+}$, with $r(Sr^{2+})$ = 1.27Å, do not fit. Therefore we assume that the expansion is due to a neighbouring $Ti^{4+}$ vacancy and is caused by the repulsion of the $O^{2-}$ ions.

**Critical effects**

*Static critical exponents*

The order parameter (with critical exponent β) corresponds to the displacement parameter, which in STO is represented by the rotation angle φ. The shaping of the crystals was such that after rapid cooling the crystals became monodomain below the structural phase transition (Fig. 2). In monodomain crystals only the ±φ$^*$ lines are present.[14]

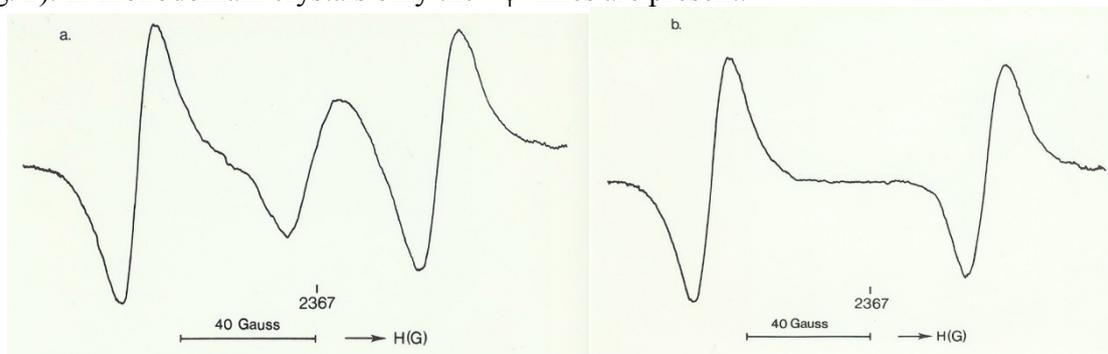

*Fig. 2. X-band EPR spectrum of the -½↔+½ transition near $g^{eff} \approx 2.8$ of a (a) three- and (b) a monodomain crystal of the non-cubic $Fe^{5+}$ in STO.*

In a monodomain crystal the rotation angle φ$^*$ can be used for the study of static critical exponents and will not be disturbed by extra EPR lines stemming from different domains. In the temperature range 30 K < $T$ < 85 K this rotation angle was found to be linearly proportional to the intrinsic one. The classical Landau behaviour with β = ½ was obtained by plotting $\varphi^{*1/\beta} = \varphi^{*2}$ as function of the reduced temperature $t = T/T_c$. For 0.7 < $t$ < 0.9 we found a straight line following a Landau behaviour. At $t$ = 0.9 the bending down from a straight line becomes noticeable (Fig. 3).

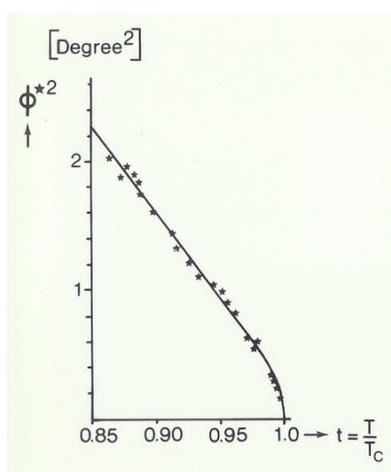

*Fig. 3. Plot of $\varphi^{*2}$ of the non-cubic $Fe^{5+}$ centre in STO versus reduced temperature $t = T/T_c$, showing the changeover from classical to critical behaviour.*

It is found that for $0.9 < t < 1$ $\beta = 1/3$, i.e., in this temperature region the crystal displays critical behaviour. In Fig. 4, $\varphi^{*1/\beta} = \varphi^{*3}$ is plotted as a function of $t$. Extrapolation of the plot to $\varphi^{*3} = 0$ yields the phase transition temperature $T_c = 103$ K which in our sample is lower than the usual value of 105 K. This is due to the presence of impurities, which can alter the phase transition temperature.

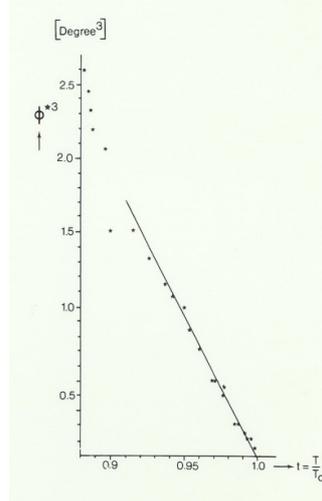

Fig. 4. Plot of $\varphi^{*3}$ of the non-cubic $Fe^{5+}$ centre in STO versus reduced temperature $t = T/T_c$.

*Asymmetric line shapes for $T \rightarrow T_c^-$*

Time dependent fluctuations and line broadening effects for the non-cubic $Fe^{5+}$ have been published before by Kool et al.[4,6] Also asymmetric line shapes in the critical region were found.[4,15,16] In a monodomain crystal STO, with the magnetic field $H\|[110]$ and the elongated axis of the crystal $c\|[001]$, we found outside the critical region ($T \ll T_c$) a symmetric Lorenzian line shape for each of the $\pm\varphi^*$ lines at $g^{eff} \approx 3.4$ On approaching $T_c$, the lines have at $T = T_c - 0.8$ K an asymmetric line shape (Fig. 5).

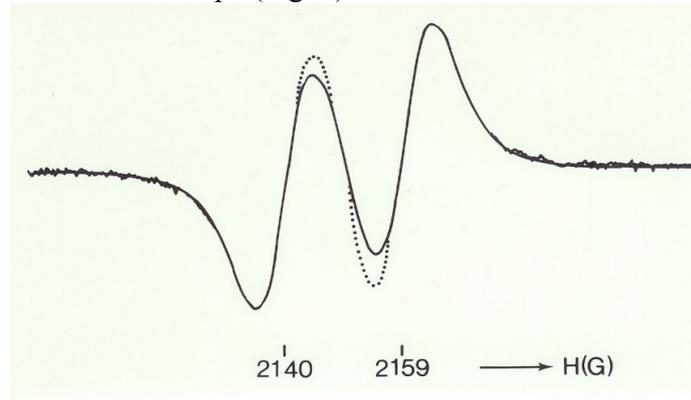

Fig. 5. Asymmetric line shape at $g^{eff} \approx 3.4$ of the non-cubic $Fe^{5+}$ centre in a monodomain crystal at X-band in the vicinity of $T_c$. The dotted line is the calculated line composed of a mixture of 50% Gaussian and 50% Lorenzian line shape. The solid curve is the experimental measured one, indicating an asymmetric line form.

The asymmetry is apparent from a comparison of the observed line shape with that of a simulated symmetric line shape. The dotted curve is simulated for a best fit symmetrical line composed of a mixture of 50% Gaussian line shape and a 50% Lorenzian line shape. The discrepancy in the amplitudes of the experimental and simulated curves is a measure of the asymmetry in the experimental line shape. At $T_c$ the fluctuations of the oxygen octahedra become very slow compared to the EPR measuring time. This means that the local rotation of each non-cubic $Fe^{5+}$ centre is seen at rest in the EPR experiment. At $T_c$ the EPR lines reflect Gaussian line shape in the case of statistical independence. The origin of the asymmetry is related to the form of the probability distribution for $\varphi^*$ near $T_c$ in the slow motion limit. Close to $T_c$, the classical probability distribution $P(\varphi^*)$ of the ensemble $P(\varphi^*,T) = c(T)\exp(-\Delta F(T))$, where $\Delta F(T)$ is the free energy depending on the order parameter. In the Landau theory $\Delta F$ is given by
$A(T)\varphi^{*2} + B\varphi^{*4}$, where $A(T) = a(T - T_c)$. $P(\varphi^*,T)$ is a double peaked function for $T < T_c$ (Fig. 6 left).

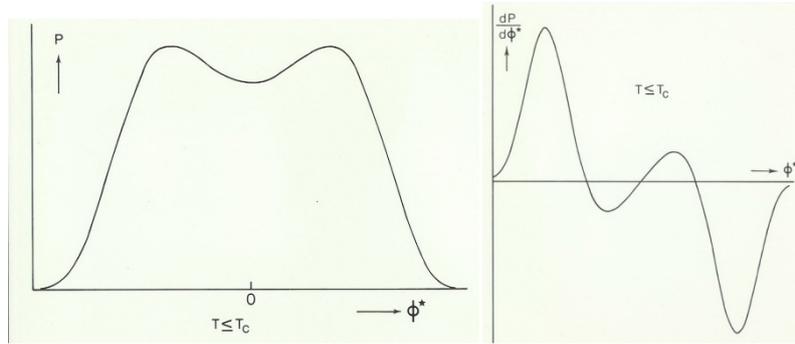

*Fig. 6. Left. Calculated probability distribution P versus the order parameter $\varphi^*$ in the vicinity of $T_c$. Right. First derivative $(dP/d\varphi^*)$ versus the order parameter.*

The spread in the value that $\varphi^*$ may adopt, of course, gives rise to an inhomogeneous broadening (spread in $g^{eff}$) of the EPR lines. In Fig. 6 (right) it is sketched how the EPR line shape is effected by the distribution function. Additional homogeneous broadening effects result in the observed asymmetrical shaped EPR lines. All the results obtained here are similar to those obtained for the $Fe^{3+}$-$V_O$ centre reflecting cooperative bulk behaviour of the crystal near the 105 K phase transition.[15,16]

**Different $Fe^{5+}$ centres**
Different $Fe^{5+}$ centres have been found in STO as well as in $BaTiO_3$ (BTO). In STO:$Fe^{5+}$, ($g_{isotropic} = 2.013$), only the $-½ \leftrightarrow +½$ transition could be observed, even at helium temperatures.[17] The $±3/2 \leftrightarrow ±1/2$ transitions are not observable due to a distribution of strain in the crystal leading to fine structure broadening. Because of the smaller radius of this centre in comparison with that of $Ti^{4+}$, the ion must be off-centred in one of the <100> directions. In contrast to this the $Fe^{5+}$ centre in BTO goes off-centre along one of the <111> directions.[18] In previous unpublished EPR investigations of STO a vanadium centre was found, which was attributed to a $V^{5+}$-$O^-$ hole like centre (Fig. 7).[19] It is a tetragonal $S = ½$, $I = 7/2$ light sensitive centre with $g_\| = 2.017$, $g_\perp = 2.012$ and $A_\| = 10.3\times10^{-4}$ cm$^{-1}$, $A_\perp = 9.9\times10^{-4}$ cm$^{-1}$.

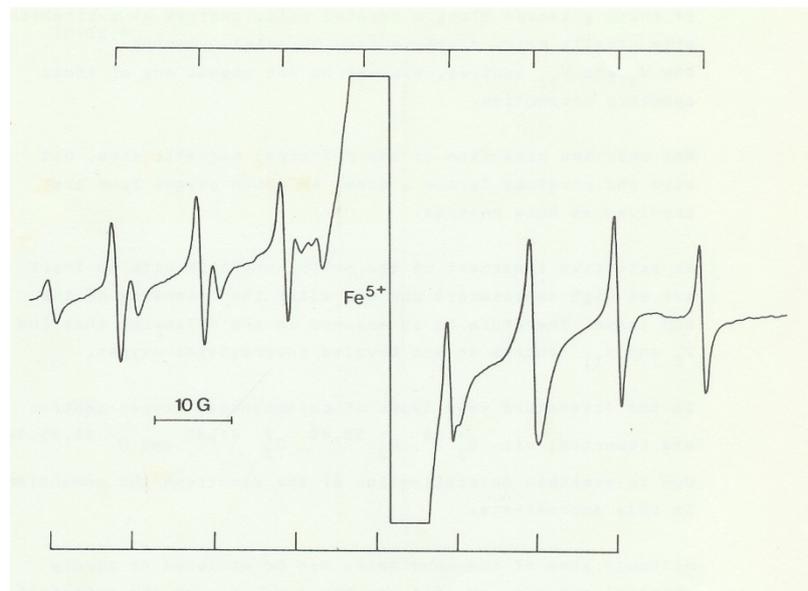

Fig. 7. X-band EPR of STO:$Fe^{5+}$-$O^{2-}$-$V^{5+}$ (reproduced from the PhD thesis of A.Lagendijk).

The interpretation of this centre was largely based on arguments used for a similar centre in STO, i.e., the $Al^{3+}$-$O^-$ hole centre.[20] In both centres the mean g-value is larger than the free electron one and the hyperfine interaction is relatively small indicating that the hole is not localized on the central ion. For instance, the localized electron in the $V^{4+}$ ($d^1$) Jahn-Teller centre in STO has much larger hyperfine values of $A_\| = 147\times10^{-4}$ cm$^{-1}$ and $A_\perp = 44\times10^{-4}$ cm$^{-1}$ [21]. However, Schirmer et al found a new $Al^{3+}$-$O^-$ hole centre in STO with a different local symmetry.[22] Therefore, they reinterpreted the formerly found hole centre to be an $Fe^{5+}$-$O^2$-$Al^{3+}$ centre, where the $Al^{3+}$ is located in the neighbouring oxygen octahedron substitutional for a $Ti^{4+}$ ion. The average g-value of this centre, $g_{av} = ⅓(g_\| + 2g_\perp) = 2.013$, is the same as the isotropic g-value found for the $Fe^{5+}$ in STO.[17] New EPR experiments revealed that by applying [011] uniaxial stress no change in the EPR line intensity took place, indicating that no reorientation of the axes of this centre occurs.[23] Also the vanadium hole centre could not be reoriented by applying uniaxial [011] stress.[23] Former stress studies showed that under influence of uniaxial externally applied stress the hole like centres $Fe^{2+}$-$O^-$ in STO[24] and $Na^+$-$O^-$ in BTO[25] could be reoriented. Therefore we ascribe this vanadium hole centre to an $Fe^{5+}$-$O^{2-}$-$V^{5+}$ association with a similar structure as the $Fe^{5+}$-$O^{2-}$-$Al^{3+}$ centre. The $g_{av} = ⅓(g_\| + 2g_\perp) = 2.0137$ is equal to that of the $Fe^{5+}$ centre in STO.[17]

All the discussed iron impurity centres are non-centro symmetric systems in contrast to, for instance, the STO:$Mo^{3+}$ [26] and STO:$Cr^{3+}$ [4] centres with $Cr^{3+}$ and $Mo^{3+}$ sitting on-centred.

**Conclusion**

The non-cubic $Fe^{5+}$ centre in STO shows the same behaviour in the vicinity of the 105 K structural phase transition as the $Fe^{3+}$-$V_O$ and $Fe^{3+}$ centres, indicating cooperative effects in the $SrTiO_3$ crystal. Furthermore the non-cubic $Fe^{5+}$ is probably an $Fe^{5+}$-$O^{2-}$-$V_{Ti}$ centre with an expanded adjacent oxygen octahedron due to a $Ti^{4+}$ vacancy. New stress experiments confirmed that the hole centre $Al^{3+}$-$O^-$ must be an $Fe^{5+}$-$O^{2-}$-$Al^{3+}$ centre and a previous found vanadium hole centre is now attributed to an $Fe^{5+}$-$O^{2-}$-$V^{5+}$ association.